\documentclass[%
 reprint,
 amsmath,amssymb,
 aps,
pra,
floatfix,
]{revtex4-2}

\usepackage{graphicx}% Include figure files
\usepackage{dcolumn}% Align table columns on decimal point
\usepackage{siunitx}
\usepackage{bm}% bold math
\usepackage[colorlinks=true,linkcolor=blue,citecolor=blue]{hyperref}
\usepackage{glossaries}
\usepackage{algorithm}
\usepackage{algpseudocode}
\usepackage{amsmath}
\usepackage{upgreek}
\usepackage{gensymb}
\usepackage{multirow}
\usepackage{booktabs}
\usepackage[export]{adjustbox}

\newcommand{\fmax}{F_\text{max}}
\newcommand{\fmin}{F_\text{min}}

\begin{document}

\preprint{APS/123-QED}

\title{Probing Strong-Field QED via Angle-Discriminated Emissions from Electrons Traversing 
Colliding Laser Pulses}

\author{Christoffer Olofsson}
\affiliation{Department of Physics, University of Gothenburg, SE-41296 Gothenburg, Sweden}
\author{Arkady Gonoskov}%
\affiliation{Department of Physics, University of Gothenburg, SE-41296 Gothenburg, Sweden}

\date{\today}

\begin{abstract}
Future laser-electron colliders will reach quantum parameters $\chi$ well in excess of unity, enabling studies of strong-field QED in extreme regimes. However, statistical inference in such experiments requires mitigating premature radiative losses of electrons to enable high-$\chi$ QED events, as well as separating the detectable signal of these events from that of lower-$\chi$ particles and photons produced by QED cascades. We propose a collider geometry in which electrons traverse the waist of two or four perpendicularly propagating, tightly focused laser pulses. This configuration suppresses both outlined difficulties by leveraging the short interaction length of the waist, rather than relying on the more technically demanding reduction of pulse duration. Moreover, altering the phase and polarization of each pulse causes the electrons to undergo helical motion where the deflection angle is correlated with the field strength, permitting an angle-based discrimination of the signal from high-$\chi$ events. Analysis and simulations show that the case of four circularly polarized pulses uniquely permits achieving helical motion throughout the entire focal region, leading to near-perfect high-$\chi$ angle-discrimination and thereby high signal-to-noise ratio. These findings support the consideration of the proposed concept as a viable layout for future experiments at PW laser facilities.
\end{abstract}

\maketitle

\section{Introduction}
With an increasing number of petawatt-class~\cite{Danson2019, Danson2004, Nakamura2017, Burdonov2021, Zhang2020} and multi-petawatt-class~\cite{Gales2018, Yoon2021, Azechi2009} laser facilities becoming operational, with several more under development~\cite{Zou2015, Li2018, Hernandez, Shen2018}, the extreme intensities achieved through tight focusing further enable applications such as relativistic particle acceleration~\cite{Qiu2025, Ziegler2024, Emmanuel2005, Borghesi2006, Zepf2003} and high-energy photon generation~\cite{Albert2016, Nerush2014, Edwards2002, Perry1999}. One ultimate objective can be the observation of Schwinger pair production, as predicted by QED in electromagnetic fields approaching the critical strength \( E_\text{cr} = m_e^2 c^3 / e \hbar \approx \SI{e+18}{\volt\per\meter} \), where \( m_e \) and \( e \) are the electron mass and charge, respectively, \( c \) is the speed of light, and \( \hbar \) is the reduced Planck constant. Producing field strengths of this order remains beyond current capabilities, but may become possible via frequency upshifting through interactions with overdense plasmas~\cite{Marklund2023, bulanov2003light, lichters1996short, baumann2019probing, vincenti2019achieving}.

Another possibility enabled by current laser technology is observing strong-field QED (SFQED) events using high-energy electrons or photons \cite{Bragin2017, Turcu2019, macleod2023strong, dinu2020trident, kettle2021laser, titov2020non}. If a high-energy electron, with Lorentz factor $\gamma$, enters a strong electromagnetic field, the observed electric field in the electron's rest frame may exceed $E_\text{cr}$. In this case, processes are described by SFQED, quantified by the electron proper acceleration
 \begin{equation} \label{eq:chi_definition}
    \chi = \gamma E_\text{cr}^{-1} \sqrt{\left(\mathbf{E} + \bm{\beta} \times \mathbf{B}\right)^2 - \left(\mathbf{E} \cdot \bm{\beta} \right)^2}
\end{equation}
where $\bm{\beta}$ is the electron velocity normalized to $c$ and $\mathbf{E}, \mathbf{B}$ are the electromagnetic field vectors. 

Experimental observations of SFQED in the regime $\chi \lesssim 1$, employing laser–electron collisions, date back to the 1990s, when both nonlinear Compton scattering (NCS) and nonlinear Breit–Wheeler (NBW) pair production were observed at SLAC \cite{Burke1997, Bula1996}. In an all-optical configuration, NCS was later observed at the Rutherford Appleton Laboratory \cite{Poder2018, Cole2018, los2024observation} and more recently at the Center for Relativistic Laser Science (CoReLS) \cite{Mirzaie2024}. In addition to laser–electron colliders, SFQED processes can also be studied using beam–plasma interactions \cite{Matheron2023}, beam–beam collisions \cite{Yakimenko2019}, and aligned crystals \cite{Wistisen2019, piazzanonperturbative, Kimball1985}.

One direction for future experiments is to explore regimes beyond $\chi \sim 1$ to induce more pronounced nonlinear effects of SFQED, including the observation of QED cascades \cite{Jirka2017} ($\chi \gtrsim 1$), higher-order processes \cite{Gonoskov2022}, radiative corrections to the electron mass \cite{Yakimenko2019}, and the conjectured breakdown of perturbative SFQED at $\chi \sim \alpha_f^{-3/2}$ \cite{ritus1972radiative, narozhny1980expansion}, where $\alpha_f \approx 1/137$ is the fine-structure constant. However, the probabilistic nature of SFQED events, combined with spatial field variations along particle trajectories and shot-to-shot fluctuations in experimental conditions, poses significant challenges for extracting quantitative data suitable for comparison with theoretical predictions.

In principle, a Bayesian framework provides an approach for handling probabilistic processes even in the presence of latent variables, such as the impact parameter between the laser and electron beam \cite{olofsson2023prospects, los2024observation}. Nonetheless, the choice of collision configuration remains crucial for invoking high-$\chi$ events while simultaneously enabling statistical inference on them. Recently, it was shown that, under the assumption of high electron energies, the maximum value of $\chi$ for a given total power is achieved using so-called bidipole focusing, while the use of circular polarization enables angle-discriminated extraction of signals from high-$\chi$ events \cite{Olofsson2022}. This setup resembles a head-on collision of electrons with an optimally focused laser pulse, while still permitting the use of focusing mirrors with effective f-numbers $f\approx 0.5$ without significant performance loss.

Nevertheless, aside from experimental constraints associated with head-on collisions and tight focusing, this setup becomes inefficient at laser powers exceeding 1PW, as the onset of QED cascades over typical pulse durations obscures the signal of interest. For electron energies around 20~GeV, this effect largely precludes studies at $\chi \gtrsim 10$.

\begin{figure}[b]
    \centering
    \includegraphics[width = \columnwidth]{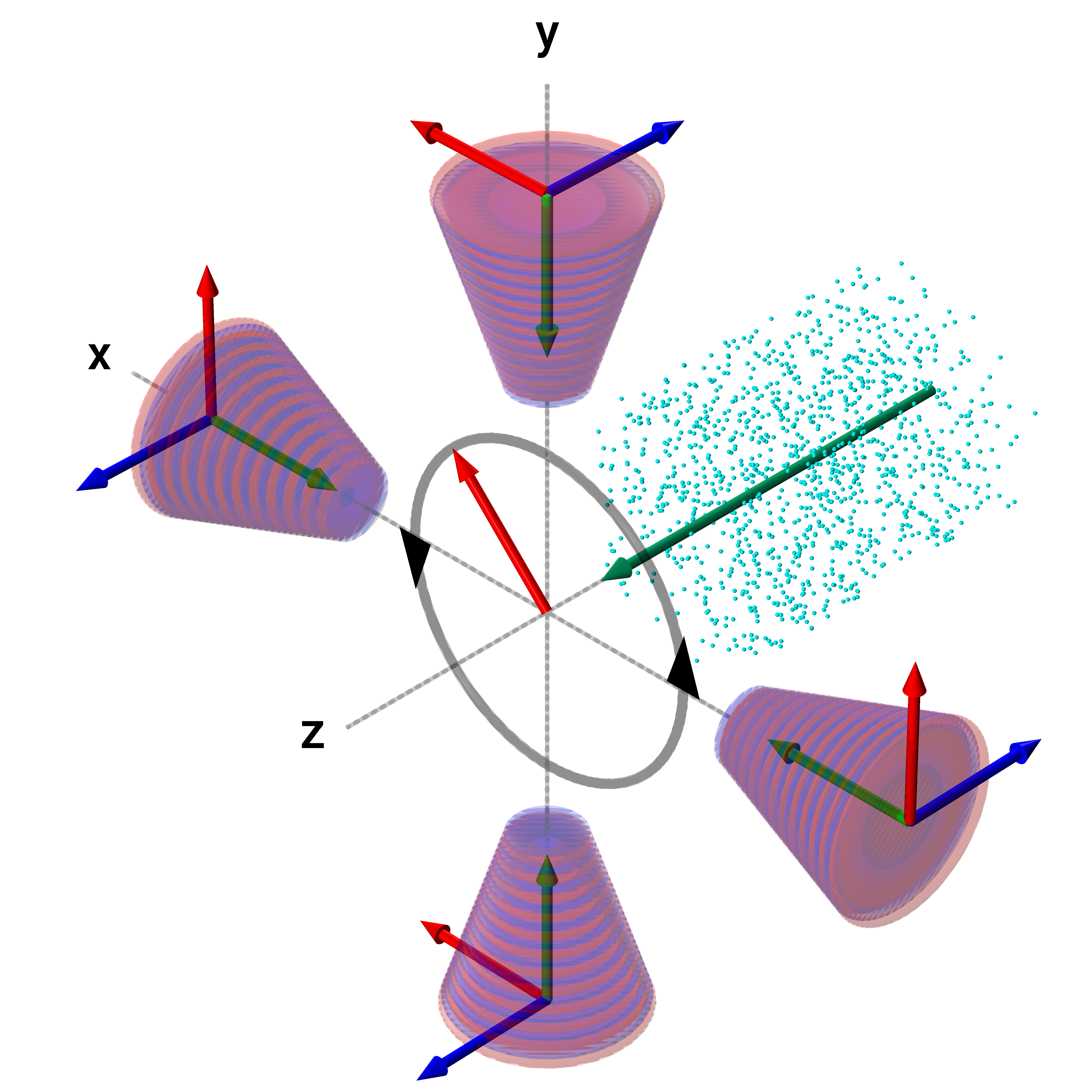}
    \caption{Four laser pulses focusing in a plane perpendicular to the electron propagation axis. Their phases are set to produce a rotating electric field (red arrow) at the focal point. Green and blue arrows denote propagation directions and the magnetic field, respectively.}
    \label{fig:crossed_pulse_setup}
\end{figure}

In this paper, we propose and assess an experimental layout that overcomes outlined difficulties, enabling the use of multi-PW laser facilities to conduct quantitative studies of SFQED in the range $\chi\sim10-100$. The layout exploits electrons propagating across the waist of tightly focused radiation to reduce the time for cascade development \cite{Gonoskov2017}, while the angle-discrimination of the signal from high-$\chi$ events is achieved by using two or four perpendicularly propagating laser pulses with properly chosen polarization and phase difference (see Fig.~\ref{fig:crossed_pulse_setup}). This layout permits various modifications, such as the use of a different number of laser pulses to increase the value of $\chi$ and the use of different polarization/phase to control the electromagnetic field observed by electrons.

\section{Problem statement} \label{sec:problem_statement}
We consider a geometry that minimizes the interaction time by requiring electrons to traverse the waist of tightly focused laser radiation. Our goal is to identify such a geometry that (i) provides high values of $\chi$ and (ii) generates a rotating electromagnetic field that induces helical electron motion which enables angle-based discrimination of signals from high-$\chi$ events. To address the first criterion, we analyze achievable $\chi$ values using multiple colliding laser pulses (MCLP) as a means to increase the peak electric field strength \cite{bulanov2010multiple}, which has been studied for $n=2,8,16$ pulses in the context of seeded QED cascades \cite{Grismayer2017, Gelfer2015, Gelfer2016}. 
We consider electrons with energy $\varepsilon \approx m_ec^2 \gamma$ and a tightly focused laser pulse of wavelength $\lambda$ and power $P$. The size of the focal spot scales as $\lambda^2$; hence, the peak electric field strength increases with intensity $I$ as $\sqrt{I} \propto \sqrt{P / \lambda^2} = \sqrt{P} / \lambda$. Consequently, $\chi$ grows as $\chi \propto \gamma \sqrt{P} / \lambda$ and under the assumptions that $|\bm{\beta}| \approx 1$ and $\gamma \gg a_{\text{max}}$ its peak value can be expressed as
\begin{equation} \label{eq:chi_max}
\chi_\text{max} = \kappa \left(\varepsilon / \si{\giga\electronvolt} \right) \left( \sqrt{P / \si{\peta\watt}} \right) \left( \lambda / \si{\micro\meter} \right)^{-1},
\end{equation}
where $a_{\text{max}}$ is the peak electric field in units of $2 \pi m_e c^2/ e \lambda$. Here, $\kappa$ quantifies how efficiently the laser power is converted into large $\chi$ values and its maximum $\kappa \approx 5.25$ occurs for bidipole focusing which implements a combination of electric and magnetic dipole pulses \cite{bassett1986limit, Gonoskov2012}, as demonstrated in Ref.~\cite{Olofsson2022}. Although $\kappa$ can be significantly lower for electrons traversing the waist of a single focused pulse, the use of $n$ such colliding pulses can partially compensate for this reduction.

This configuration directly addresses point (ii), as rotating electromagnetic fields at the focus can be realized by adjusting each pulse’s polarization and phase. As electrons traverse a region where the electromagnetic field vectors rotate around the direction of propagation, their emission angle $\alpha \sim a_{\text{max}} / \gamma$, defined as the angle between the electron's initial propagation axis and its direction at the moment of emission, becomes correlated with $\chi$ as both depend on the local field strength. This correlation allows the selection of photons emitted by high-$\chi$ electrons based on $\alpha$. Furthermore, photons of frequency $\omega$ that reach the detector with high energy $\delta = \hbar \omega / m_e c^2$ are less likely to have originated from newly formed particles or subsequent emissions, both associated with lower $\chi$ values.

In addition to suppressing a QED cascade, the outlined geometry permits the use of longer laser pulses to mitigate the effect of temporal shot-to-shot variations, which cause the overlap between the electrons and the strong-field region to fluctuate. We note that the spatial misalignment can be mitigated by allowing the electron beam to de-focus, such that the strong-field region forms within a uniform stream of electrons. While this approach reduces the number of high-$\chi$ events, we adopt this configuration as it corresponds to data accumulation over multiple shots.

In the following section we discuss the implementation of the outlined interaction layout and its impact on the value of $\kappa$. We also specify how the polarization and phase of each pulse is chosen to form rotating electromagnetic fields at the focal point to achieve the correlation between $\chi$ and $\alpha$.

\section{Interaction geometry}
Assuming linear (LP) or circular (CP) polarization for each of the $n$ colliding pulses, all having a central frequency of $\omega_0$, we examine conditions for generating a rotating electric field at focus. Analogously, a rotating magnetic field can be formed but the impact on $\kappa$ and the correlation $\alpha \propto \chi$ remain unchanged in the ultrarelativistic limit $|\bm{\beta}| \approx 1$. The resulting electric field at focus can be expressed as the real part of a complex field vector $e^{-i\omega_0 t}\mathbf{E}^{(n)}$, satisfying $\bm{\beta} \cdot \mathbf{E}^{(n)} = 0$. In the vicinity of the focal region, $\mathbf{E}^{(n)}$ may be approximated as a superposition of plane waves with frequency $\omega_0$ and wave number $k$
\begin{equation} \label{eq:gen_e}
    \mathbf{E}^{(n)} = \frac{E_0}{\sqrt{n}} \sum_{j=1}^{n} e^{i k (\hat{k}_j \cdot \mathbf{r})} \xi_j \hat{e}_j,
\end{equation}
where $E_0 \propto \sqrt{P}$ correspond to the peak electric field of a single laser pulse, $\hat{k}_j$ is the unit wave vector and $\hat{e}_j$ are complex unit vectors indicating each pulse’s electric field direction and phase. The polarization of each wave is parameterized by $\xi_j = 1$ for LP or $\xi_j = 1/\sqrt{2}$ for CP. The corresponding growth of $\kappa$, denoted as $\kappa^{(n)} \propto \chi^{(n)}_{\text{max}}$, is proportional to Eq.~\eqref{eq:chi_definition} at $\mathbf{r}=t=0$ (assuming all waves are synchronized) and is expressed using Eq.~\eqref{eq:gen_e} as
\begin{equation} \label{eq:general_kappa}
    \kappa^{(n)} \propto \frac{E_0}{\sqrt{n}} \left| \sum_{j=1}^{n} \xi_j \left( \hat{e}_j (1-\bm{\beta} \cdot \hat{k}_j) + (\bm{\beta} \cdot \hat{e}_j) \hat{k}_j \right) \right|.
\end{equation}

Using the coordinate convention of Fig.~\ref{fig:crossed_pulse_setup}, the value of $\kappa$ for the case of electrons colliding head-on with a single pulse, denoted as $\kappa^{(\text{dir})}$, is found by setting $\hat{k}_1 = -\hat{z}$ and $\bm{\beta} \approx \hat{z}$, so that
\begin{equation} \label{eq:single_pulse_kappa_scaling}
    \kappa^{(\text{dir})} \propto \xi_1 E_0.
\end{equation}
This geometry gives rise to rotating electromagnetic fields only if the pulse is circularly polarized.

\begin{table*}[ht]
\centering
\caption{Values of $\kappa$ and $\chi_\text{max}$ for various powers, electron energies, f-numbers and interaction layouts. Bold entries indicate geometries that suppress QED cascades by letting electrons traverse the waist of the intersecting pulses.}
\label{tab:kappa_scaling}
\begin{tabular}{c | c | c|c|c|c | c|c|c|c}
\toprule
\multirow{2}{*}{\(f\)} & \multirow{2}{*}{Geometry}
& \multicolumn{4}{c|}{Linear Polarization} 
& \multicolumn{4}{c}{Circular Polarization} \\
 & & \(\kappa\) & \shortstack{\(\chi_\text{max}\)\\[-0.3ex]\(\SI{1}{\peta\watt},\SI{10}{\giga\electronvolt}\)} & \shortstack{\(\chi_\text{max}\)\\[-0.3ex]\(\SI{3}{\peta\watt},\SI{20}{\giga\electronvolt}\)} & \shortstack{\(\chi_\text{max}\)\\[-0.3ex]\(\SI{10}{\peta\watt},\SI{100}{\giga\electronvolt}\)} & \(\kappa\) & \shortstack{\(\chi_\text{max}\)\\[-0.3ex]\(\SI{1}{\peta\watt},\SI{10}{\giga\electronvolt}\)} &\shortstack{\(\chi_\text{max}\)\\[-0.3ex]\(\SI{3}{\peta\watt},\SI{20}{\giga\electronvolt}\)} & \shortstack{\(\chi_\text{max}\)\\[-0.3ex]\(\SI{10}{\peta\watt},\SI{100}{\giga\electronvolt}\)} \\
\midrule
 & Head-on & 2.04 & $\sim 25$ & $\sim 88$ & $\sim 806$ & 1.44 & $\sim 18$ & $\sim 62$ & $\sim 570$ \\
1.0 & $\mathbf{n=2}$ & \textbf{0.72} & $\sim \mathbf{9}$ & $\sim \mathbf{31}$ & $\sim \mathbf{285}$ & $-$ & $-$ & $-$ \\
 & $\mathbf{n=4}$ & \textbf{1.02} & $\sim \mathbf{13}$ & $\sim \mathbf{44}$ & $\sim \mathbf{403}$ & \textbf{1.44} & $\sim \mathbf{18}$ & $\sim \mathbf{62}$ & $\sim \mathbf{570}$ \\
\midrule
 & Head-on & 2.53 &  $\sim 31$ & $\sim 109$ & $\sim 1000$ & 1.79 & $\sim 22$ & $\sim 77$ & $\sim 707$ \\
0.75 & $\mathbf{n=2}$ & \textbf{0.89}  & $\sim \mathbf{11}$ & $\sim \mathbf{38}$ & $\sim \mathbf{351}$ & $-$ & $-$ & $-$ \\
 & $\mathbf{n=4}$ & \textbf{1.27} & $\sim \mathbf{16}$ & $\sim \mathbf{55}$ & $\sim \mathbf{502}$ & \textbf{1.79} & $\sim \mathbf{22}$ & $\sim \mathbf{77}$ & $\sim \mathbf{707}$ \\
 \midrule
 & Head-on & 3.22 & $\sim 40$ & $\sim 139$ & $\sim 1272$ & 2.28 & $\sim 28$ & $\sim 98$ & $\sim 900$ \\
0.5 & $\mathbf{n=2}$ & \textbf{1.14} & $\sim \mathbf{14}$ & $\sim \mathbf{49}$ & $\sim \mathbf{450}$ & $-$ & $-$ & $-$ \\
 & $\mathbf{n=4}$ & \textbf{1.62} & $\sim \mathbf{20}$ & $\sim \mathbf{70}$ & $\sim \mathbf{640}$ & \textbf{2.28}  & $\sim \mathbf{28}$ & $\sim \mathbf{98}$ & $\sim \mathbf{900}$\\
\midrule
$\lesssim 0.25$ & Bidipole & $5.25$ & $\sim 65$ & $\sim 227$ & $\sim 2075$ & $3.71$ & $\sim 46$ & $\sim 160$ & $\sim 1467$ \\
\bottomrule
\end{tabular}
\end{table*}
Proceeding to the case of $n=2$ pulses, the only possible configuration is to employ pulses with linear polarization. Here, a rotating electric field emerges when one pulse arrives from the x-axis (see Fig.~\ref{fig:crossed_pulse_setup}), providing an electric field component perpendicular to that of the second pulse, arriving from the y-axis with a relative phase difference of $\pi/2$. Consequently, the associated magnetic field satisfies $\bm{\beta} \times \mathbf{B} = 0$ (or $\mathbf{B}=0$ for a standing wave when the phase difference is zero). The corresponding value of $\kappa^{(2)}$ as compared to Eq.~\eqref{eq:single_pulse_kappa_scaling} is found from Eq.~\eqref{eq:general_kappa} by setting e.g. $\hat{k}_1 = \hat{y}$, $\hat{k}_2=\hat{x}$, $\hat{e}_1 = i\hat{x}$ and $\hat{e}_2 = \hat{y}$
\begin{equation} \label{eq:kappa_2_scaling}
    \kappa^{(2)} = \frac{\kappa^{(\text{dir})}}{2\sqrt{2}\xi_1}.
\end{equation}
Notably, this particular setup ensures that in an experiment, neither the electron beam nor the laser pulses impinge on any critical instrumentation positioned along their respective axes. 
\begin{figure}
    \centering
    \includegraphics[width = \columnwidth]{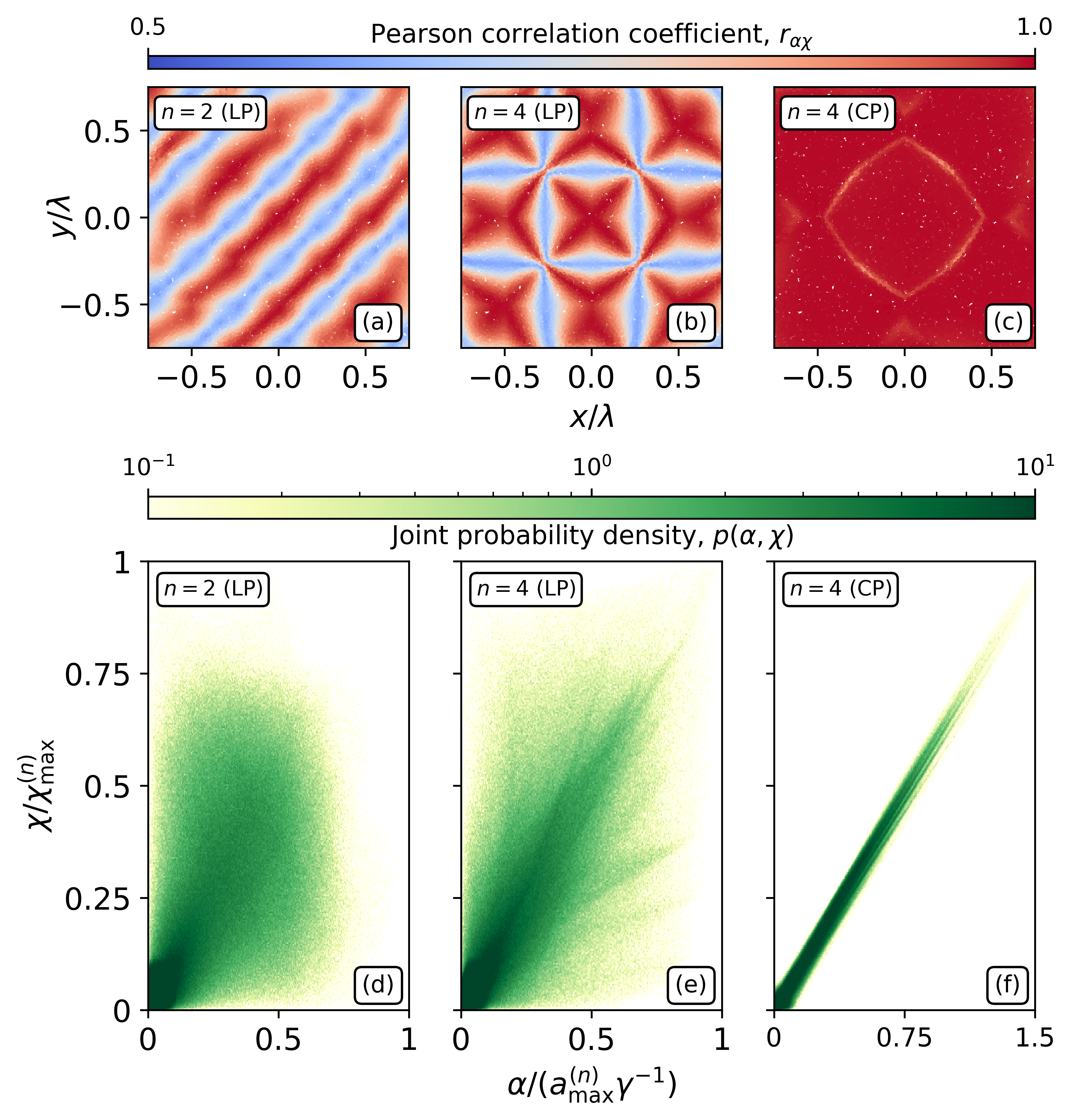}
    \caption{Correlation strength between $\chi$ and $\alpha$ computed from electron trajectories as a function of their initial transverse location (a)-(c) and the corresponding joint probability density for the electron $\chi$ and $\alpha$ values (d)-(f).}
    \label{fig:spatial_interference}
\end{figure}

Electrons whose trajectories pass through $x = y = 0$ will experience a circularly rotating electric field, establishing the $\alpha$-$\chi$ correlation described in the preceding section. Around the strong-field region, the strength of this correlation varies spatially due to interference between the two waves. When $\mathbf{r}\neq0$,  electrons encounter the electric field given by Eq.~\eqref{eq:gen_e} and is proportional to
\begin{equation} \label{eq:n2_interference}
    \propto \hat{x} \sin(ky-\omega_0 t) + \hat{y} \cos(kx-\omega_0 t).
\end{equation}
The manifest spatial phase difference in Eq.~\eqref{eq:n2_interference} leads to an elliptically rotating or even standing electric field, thereby reducing or eliminating the correlation strength. To quantify this spatial dependence, we introduce the ellipticity 
\begin{equation}
    \epsilon = 1 - \rho,
\end{equation}
as a measure of the shape traced by the tip of the electromagnetic field component responsible for transverse electron acceleration. Here,
\begin{equation}
    \rho = \frac{\fmax - \fmin}{\fmax + \fmin}
\end{equation}
is a ratio containing the maximum $\fmax$ and minimum $\fmin$ of the squared field strength over one optical cycle. Circular rotation is demarcated by $\fmax = \fmin$ whereas a standing wave is characterized by $\fmin = 0$, yielding no correlation. Whenever $\fmax = \fmin = 0$, we define $\epsilon=0$ as electrons pass through unaffected. Using Eq.~\eqref{eq:n2_interference} as input, the ellipticity for the present configuration becomes
\begin{equation}
    \epsilon = 1-\left| \sin \left(k(y-x)\right) \right|,
\end{equation}
and the strongest correlation occurs along lines $y=x+m \cdot \frac{\lambda}{2}$, where $m$ is an integer. This expression agrees with the spatial structure of the $\alpha$-$\chi$ correlation observed in simulated electron trajectories shown in Fig.~\ref{fig:spatial_interference}(a).

Moving on to the case of $n=4$ pulses, two possible configurations exists; one for which all beams have LP and the other CP. Furthermore, the pulses need to be paired in the counter-propagating arrangement shown by Fig.~\ref{fig:crossed_pulse_setup}, where one pair has perpendicular propagation axis to that of the second pair, which has an additional phase difference of $\pi/2$. For LP, the rotating electric field is similar to the previous case except that at focus the electric field contribution is doubled and $\mathbf{B}=0$. For instance, setting $\hat{k}_1=-\hat{k}_2 = \hat{x}$, $\hat{k}_3 = -\hat{k}_4 = \hat{y}$ and $\hat{e}_1 = \hat{e}_2 = \hat{y}$, $\hat{e}_3 = \hat{e}_4 = i\hat{x}$ in Eq.~\eqref{eq:general_kappa} we find that $\kappa^{(4)}$ is related to $\kappa^{(\text{dir})}$ by
\begin{equation} \label{eq:kappa_4}
    \kappa^{(4)} = \frac{\kappa^{(\text{dir})}}{2\xi_1} \ (\text{LP}),
\end{equation}
and that the corresponding electric field is proportional to
\begin{equation} \label{eq:n4_Efield}
   \propto \hat{x} \sin(\omega_0 t) \cos(ky) + \hat{y} \cos(\omega_0 t) \cos(kx). 
\end{equation}
The ellipticity that results from Eq.~\eqref{eq:n4_Efield} can be expressed as
\begin{equation}
    \epsilon = \frac{2\cos^2(kx)\cos^2(ky)}{\cos^4(kx) + \cos^4(ky)},
\end{equation}
where $\epsilon=1$ is now achieved when $y = \pm x + m \cdot \frac{\lambda}{2}$ and its spatial dependence is consistent with that found in Fig.~\ref{fig:spatial_interference}(b).

When all beams are circularly polarized, a rotating electric field arises only if each pair of counter-propagating pulses have opposite helicity. Now, let $\hat{e}_{1,2} = (\hat{y} \pm i \hat{z})$ and $\hat{e}_{3,4} = i(\hat{x}\pm i\hat{z})$ so that the scaling of $\kappa^{(4)}$ relative to $\kappa^{(\text{dir})}$ becomes
\begin{equation} \label{eq:kappa_CP_kappa_dir}
    \kappa^{(4)} = \frac{\kappa^{(\text{dir})}}{\sqrt{2} \xi_1} \ (\text{CP}). 
\end{equation}
Numerical values of $\kappa$ for several f-numbers are listed in Table~\ref{tab:kappa_scaling}, confirming the predicted scalings. Notably, electrons may also experience a nonzero magnetic field $\mathbf{B}^{(4)}$ such that they are accelerated by the component $\mathbf{E}^{(4)} + \hat{z}\times \mathbf{B}^{(4)}$ which is proportional to
\begin{equation} \label{eq:n4CP_EB}
\begin{split}
  \propto& \left(\cos(kx)+\cos(ky)\right) \cdot \left(\hat{x}\sin(\omega_0 t) + \hat{y}\cos(\omega_0 t)\right) \\
    &+ \hat{z}\left(\cos(\omega_0 t)\sin(kx) + \sin(\omega_0 t)\sin(ky)\right).
\end{split}
\end{equation}
Evidently, the transverse component of Eq.~\eqref{eq:n4CP_EB} is circularly rotating everywhere in space. This means that $\epsilon$ is always unity (see Fig.~\ref{fig:spatial_interference}(c)) except when the electromagnetic fields vanishing along $y=\pm x+m \cdot \frac{\lambda}{2}$ where now $(m \neq 0)$. Consequently, the overall correlation strength is the largest for this geometry and becomes weaker for $n=2$ and $n=4$ (LP). This is further confirmed by calculating the joint probability density $p(\alpha, \chi)$ from all simulated electron trajectories, displayed in Fig.~\ref{fig:spatial_interference}(d)-(f).

Generalization beyond $n=4$ is not treated in detail here, except to estimate an upper bound on the number of pulses. To prevent beam overlap outside the focal region, and assuming each pulse is confined within a cone of aperture angle $\theta = \arctan{\frac{1}{2f}}$ the condition $n \theta \leq \pi$ must be satisfied. For the geometries considered here, where beams propagate along orthogonal axes, this yields a lower bound on the f-number: $f\geq0.5$.

In the following section, we compare the effectiveness for geometries generating rotating fields via $n=2$ (LP) and $n=4$ (CP) pulses with a single, directly colliding laser pulse with circular polarization where $\kappa^{(4)}=\kappa^{(\text{dir})}$ holds.

\section{Numerical analysis}
The effectiveness of each interaction geometry is assessed by simulating the number of signal photons produced as a function of laser power. We introduce the signal-to-noise ratio $\eta = N_{\text{signal}}/N_{\text{total}}$ as outlined in Ref.~\cite{Olofsson2022}, where $N_{\text{total}}$ is the number of detected photons satisfying $\delta \geq 0.5$ and $\alpha \geq c_{\alpha} a_\text{max} \gamma^{-1}$. The parameter $c_{\alpha}$ represents the effective size of a spatial filter that selects off-axis photons. The number of signal photons, $N_{\text{signal}} \leq N_\text{total}$, are those which were also emitted at $\chi \geq \chi_{\text{max}}/2$ by particles retaining at least $99 \%$ of their initial energy. Simulations were performed using the energy-conserving, dispersion-free particle-in-cell code $\pi$-PIC \cite{Gonoskov2024}, extended to include NCS, NBW, QED cascades, and higher-order processes based on the implementation described in Ref.~\cite{gonoskov2015extended}.

All laser pulses are assigned a central wavelength of $\lambda = \SI{800}{\nano\meter}$ and a pulse duration of $\tau = \SI{30}{\femto\second}$. Each pulse is assumed to be focused by a mirror with $f = 1$, and modeled in the far field as occupying a spherical sector. The spatial beam profile follows the prescription detailed in Sec.~2.2 of Ref.\cite{panova2021optimized}, with the pulse center initially located at a distance $R_0 = 30\lambda$ from the origin. 

The electron beam has a density of $n_e = \SI{e19}{\per\cubic\centi\metre}$ and is represented of $\SI{2e5}{}$ macroparticles, uniformly distributed within a cylindrical volume of radius $r_e = 0.9 \lambda$ and length $\ell_e = 1.5 \lambda$ respectively. The electron energy is set to $\varepsilon = \SI{20}{\giga\electronvolt}$, which can be achieved either using an all-optical setup based on laser wakefield acceleration \cite{kim2021multi}, or through conventional accelerator technologies \cite{yabashi2015overview, tschentscher2017photon, emma2010first}. 

\begin{figure}
    \centering
    \includegraphics[width = \columnwidth]{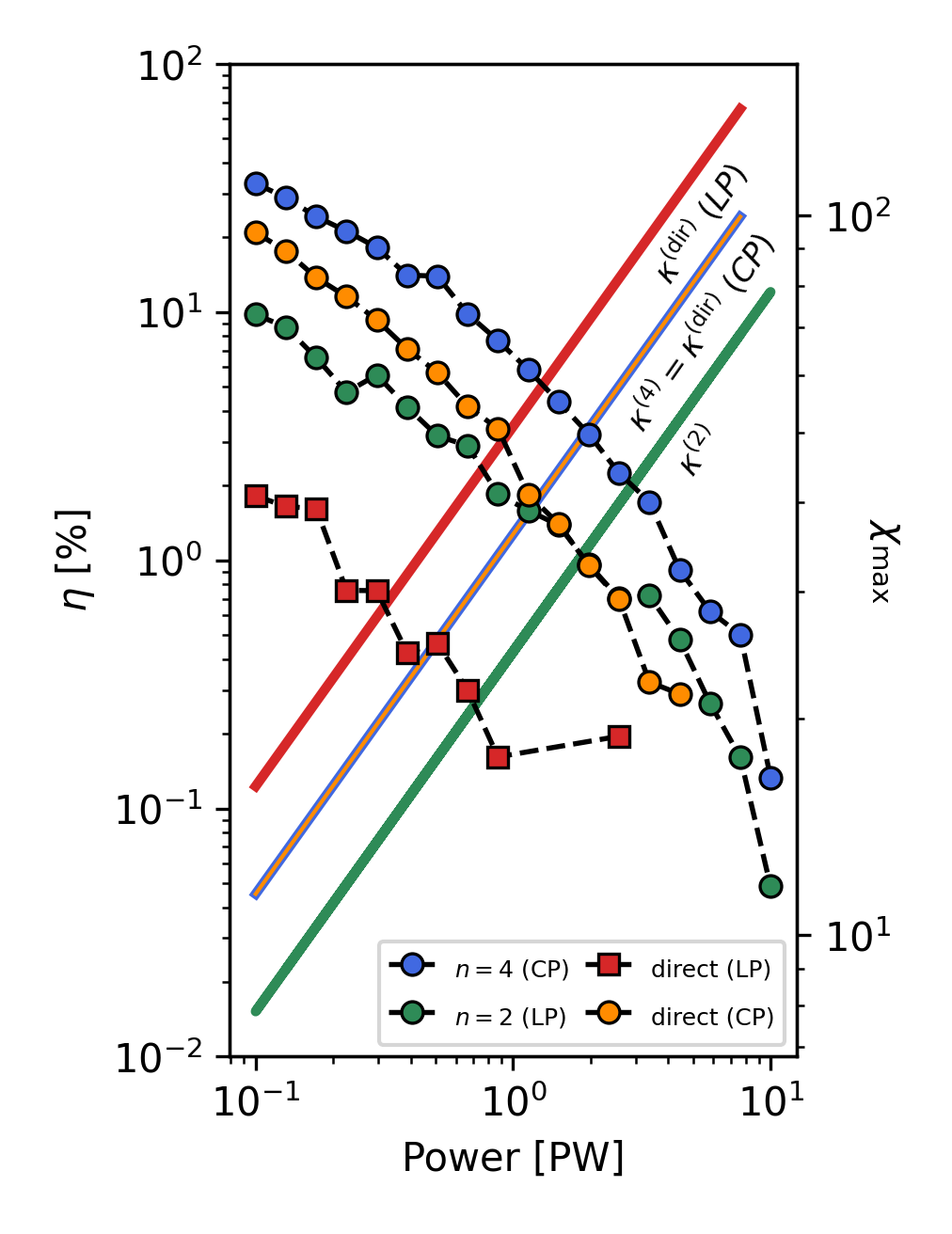}
    \caption{Signal-to-noise ratio (markers) and $\chi_\text{max}$ (solid lines) as a function of laser power. Points where $\eta = 0$ are omitted.}
    \label{fig:geometry_comparison}
\end{figure}

The computational domain is a cubic grid consisting of $128^3$ cells with a side length of $32 \lambda$. Initially, each pulse is propagated so that it is located $L = 10 \lambda$ away from the origin, which also denotes the distance to the center of the electron beam. The simulation is then carried out for a duration of $2L/c$, with a time step of $\Delta t = \frac{\lambda}{32c}$. At the end of the simulation, $\eta$ is evaluated using $c_{\alpha} = 0.6$ except for the case of $n=4$ (CP) where $c_{\alpha} = 1.0$, and the resulting parameter scan is presented in Fig.~\ref{fig:geometry_comparison}.

When $n = 2$ (LP) pulses are used, the signal-to-noise ratio reaches approximately $1\%$ at an input power of $\SI{2}{\peta\watt}$, corresponding to $\chi_\text{max} \sim 30$. We emphasize that in this configuration, neither the electron beam nor the laser pulses propagate toward sensitive equipment; however, this advantage comes at the cost of lower $\chi_\text{max}$ values compared to direct collisions. 

For directly colliding pulses with circular polarization, the signal-to-noise ratio is slightly higher at sub-petawatt powers but eventually declines due to the onset of QED cascades which arise from the relatively long interaction time. In contrast, while linearly polarized pulses yield higher $\chi_\text{max}$ values, the absence of a strong $\alpha$–$\chi$ correlation significantly reduces $\eta$. 

Finally, the configuration with $n = 4$ (CP) pulses yields $\chi_\text{max}$ values identical to those of the direct collision geometry, as predicted by Eq.~\eqref{eq:kappa_CP_kappa_dir}. Nevertheless, this setup achieves the highest signal-to-noise ratio among all configurations, reaching $\eta \lesssim 1\%$ at $\chi_\text{max} \sim 100$.
This is a result of the strong correlation inherent to this geometry, which increases the likelihood of identifying a high-$\chi$ photon based on its emission angle. To verify this behavior, we performed additional simulations without quantum processes for all considered interaction geometries at a fixed power of $\SI{1}{\peta\watt}$. For each electron, the values of $\alpha$ and $\chi$ were evaluated at each timestep to estimate their joint probability density $p(\alpha, \chi)$, shown in Fig.~\ref{fig:spatial_interference}. A similar correlation is obtained for directly colliding pulses (see Fig. 2 of Ref. \cite{Olofsson2022}). 

In the absence of electrons, the case $n=4$ (LP) made to generate a rotating, purely magnetic field ($\mathbf{E}=0$) provides a basis to study nonlinear QED driven by virtual electron loops \cite{Marklund2023}. These nonlinearities are dictated by the Heisenberg-Euler interaction Lagrangian and for supercritical magnetic field strengths, higher-order corrections increase logarithmically \cite{karbstein2019all}, eventually requiring re-summation which precipitates the creation of ultrastrong magnetic fields via high-intensity lasers.

In conclusion, $\si{\peta\watt}$-class laser systems combined with multi-$\si{\giga\electronvolt}$ electron beams can achieve $\chi_\text{max} \sim 10$–$60$ for the $n = 2$ and $n = 4$ pulse configurations (see Table~\ref{tab:kappa_scaling}) with the opportunity to mitigate a QED cascade and retrieve a SFQED signal from photons. Assuming a laser power of $\SI{10}{\peta\watt}$ and projected advances toward $\SI{100}{\giga\electronvolt}$ electron accelerators \cite{Nakajima2014}, values of $\chi_\text{max} \sim 285$–$570$ become accessible. Moreover, by considering the corresponding values of $\kappa^{(n)}$, the use of a plasma converter for the colliding pulses \cite{Marklund2023} could potentially raise $\chi_\text{max}$ to $\sim \SI{3.7e+4}{}$ for the $n = 2$ (LP) setup and $\sim \SI{7.4e+4}{}$ for the $n = 4$ (CP) configuration. 

Lastly, even if $\eta \ll 1\%$ and spatiotemporal variations are present, it is still possible to test theoretical predictions of SFQED by accumulating data from many repeated collisions and employ Bayesian techniques, such as approximate Bayesian computation \cite{olofsson2023prospects}.

\section{Conclusions}
In this paper we have elaborated and assessed a laser-electron collider geometry where electrons are delivered through the waist of either $n=2$ or $n=4$ colliding laser pulses that collectively forms a rotating electromagnetic field at focus. This enables the separation of photons originating from the strong-field region based on their emission angle, while minimizing the electron interaction time and thereby suppressing the development of significant QED cascades and their associated low-$\chi$ contributions. Our results indicate that dividing the available power between two laser pulses yields a signal-to-noise ratio comparable to that of a single head-on circularly polarized pulse, while avoiding irradiation of optical components and the electron source. By considering four such pulses with circular polarization, the overall values of $\chi$ and signal-to-noise ratio are increased. This suggests that multi-$\SI{}{\peta\watt}$ laser systems could access and probe SFQED processes where $\chi \sim 10 - 100$. At even higher powers, it may become necessary to increase the electron energy to further prevent the major impact of a QED cascade. Additionally, the use of repeated experiments in conjunction with Bayesian methods to draw inferences despite uncontrollable variations could be particularly relevant when $\eta \ll 1 \%$ \cite{olofsson2023prospects}. 

\section{Acknowledgments}
The authors acknowledge support from the Swedish Research Council (Grant No. 2017-05148) as well as computational resources provided by the Swedish National Infrastructure for Computing (SNIC).
\bibliography{literature}
\end{document}